\documentstyle[sprocl]{article}   
\input epsf
\epsfverbosetrue

\begin{document}

\title{String breaking on a lattice}

\author{P. Pennanen}

\address{Nordita, Blegdamsvej 17, DK-2100 Copenhagen \O, Denmark\\
E-mail: {\tt petrus@nordita.dk}}

\maketitle\abstracts{
String breaking is a non-perturbative long-distance feature of QCD that is 
involved in for example meson decays. A mixing analysis of lattice operators 
at zero temperature gives the level splitting and mixing energy between
the broken and unbroken string states.  }

\section{Background}

A major motivation for studying string breaking in QCD has been its
nature as a fundamental feature, sometimes even taught to high-school
students, that hadn't been reproduced from the theory. A phenomenological
motivation is its significant role in decays of e.g. mesons. The problems 
in a theoretical reproduction have on the analytical side been the 
non-perturbative nature of string breaking, while on the non-perturbative
side the long distance range has hindered lattice methods.

An analogue for $q\bar{q}$ creation in a chromoelectric QCD flux tube 
is the creation of real $e^+e^-$ in a constant electric QED field. The 
probability of the latter was found to be $\propto \exp{-\pi m^2/|eE|}$ by
Schwinger~\cite{sch:51}. A similar factor of $\exp{-4m^2 / b_s}$ for a flux 
tube with string tension $b_s$ has been calculated in strong coupling 
QCD~\cite{dos:86}. In other words, the virtual particle and anti-particle
have to separate by tunneling a distance inversely proportional to the force 
experienced in the (chromo)electric field in order to balance out the energy
needed for their masses. In vacuum the $q\bar{q}$ are produced with 
$J^{PC}=0^{++}$, which is known as the Quark Pair Creation 
model~\cite{yao:74}. For creation in a flux tube, however, only the  
component of $J$ parallel to the tube is a good quantum number.

The quark pair creation model has been combined with a flux tube model to 
calculate decay widths of hybrid mesons~\cite{isg:85b,pag:98}. Here the 
flux tube is taken to be a string of coupled quantum mechanical harmonic
oscillators vibrating in transverse planes, which predicts level orderings
well but flux tube shapes poorly~\cite{gre:96}. With an 
$L$-dependent $q\bar{q}$ creation vertex a selection rule suppressing 
decays of low-lying hybrids to identical mesons was obtained~\cite{pag:98},
which is relevant to our lattice studies. 
%In this model the excited flux 
%tube de-excites or breaks locally, which is favorable in terms of the 
%overlap of the wavefunctions. Creation of the light quark next to the heavy
%antiquark $\bar{Q}$ has also been preferred~\cite{dru:98b} as the energy
%in the tube is then cancelled. 

\section{Lattice operators}

%As shown in Fig.~1, 
The $Q\bar{Q}$ ground state potential $V_0(R)$ crosses 
twice the energy of a
$Q\bar{q}$ meson at approximately 1.2 fm in both quenched and unquenched
lattice QCD. The parameters of these simulations can be found in 
Ref.~\cite{pen:99}. Even though one might expect the Wilson loop measuring
$V_0(R)$ to level out on unquenched lattices this hasn't been seen despite
extensive efforts (for a review see~\cite{sch:99}). The reason seems to
be a poor overlap of the Wilson loop with the $Q\bar{q}+\bar{Q}q$ 
state which gets significant only at temperatures close to the 
deconfinement transition (see~\cite{sch:99,pen:00} for references). 

A variational approach, where the operators representing both the $Q\bar{Q}$ 
and $Q\bar{q}+\bar{Q}q$ states are explicitly present, has shown much more
promise with Higgs and adjoint colour source 
calculations. Here a matrix $C(R,T)$ of 
the correlations between these operators is diagonalised according to
\begin{equation}
C(R,T)v_\alpha(T) = \lambda_\alpha(T) C(R,T-1) v_\alpha(T). \label{evari}
\end{equation}
This gives eigenvalues $\lambda=\exp{-E(R)}$ and eigenvectors $v$ for each 
state $\alpha$.

In the simplest case of a $2\times 2$ matrix the eigenenergies $E_\alpha(R)$ 
can be obtained by diagonalising
\begin{equation}
\left(\begin{tabular}{cc} $V_0(R)$ & $x(R)$ \\ $x(R)$ & $M_0(R)$ \end{tabular}\right),
\label{eematrix}
\end{equation}
where $M_0(R)$ is the ground state energy of the meson-antimeson system and 
$x(R)$ is its 
mixing energy with the $Q\bar{Q}$ state. At the string breaking point
$R=R_b$ the potential $V_0(R)=M_0(R)$ and the splitting between 
$E_{0,1}$ is $2x$.

\begin{figure}[ht]
\vspace*{-3.1cm}
\begin{center}
\hspace*{-1.0cm}\epsfxsize=450pt\epsfbox{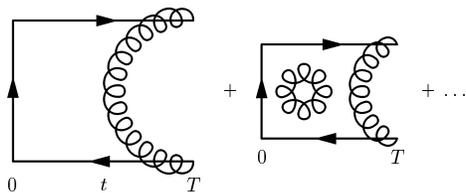}
\end{center}
\vspace*{-17.0cm}
 \caption{Illustration of Eq.~\protect\ref{ehlcq}.}
 \label{fehlc}
\end{figure}

The observables of the variational approach are also very hard to measure 
on the lattice, as the gap $x$ is small. Even with the use of all-to-all 
propagator estimates~\cite{mic:98} that give better 
statistics than conventional techniques it was hard to get a proper signal.
Fortunately we found a way to measure $x$ explicitly by extracting it from 
the lattice correlators. 

As illustrated in Fig.~1, the correlation between $Q\bar{Q}$ and 
$Q\bar{q}+\bar{Q}q$ states can be pictured as propagation of the two-quark 
state from time 0 until time $t$, where $q\bar{q}$ are created with mixing energy $x$ and propagate until time $T$. In the unquenched theory additional terms 
with more mixings come from vacuum bubbles appearing inside the diagram. This 
can be written as
\begin{equation}
U(T) = x(R) \sum_{t=0}^T  \sum_{k=0}^\infty w_k e^{-V_k(r)t}  
\sum_{l=0}^\infty e^{-M_l(r)(T-t)} d_l \  [+O(x^3)]_{\rm unquenched}. 
              \label{ehlcq} 
\end{equation}
The connected correlator $B$ of the meson-antimeson state with itself can be 
written in the same manner. When the Wilson loop $W(T)=\sum_{k=0}^\infty 
w_k^2 e^{-V_k(r)T}$ and the unconnected correlator 
$D(T)=\sum_{l=0}^\infty d_l^2 e^{-M_l(r)T}$ are measured separately the $B$ and
$U$ provide two different ways of extracting the mixing $x$~\cite{pen:00}.

We obtained a first result for its value in unquenched SU(3) lattice QCD
at the string breaking point: $x/a=46(8)$ MeV. Its small value enables the use
of only the first term in Eq.~\ref{ehlcq} and the corresponding equation for 
box, and makes it difficult to observe the mixing directly from the spectrum.
From the Schwinger factor we expect $x$ to be roughly constant for
$R\approx R_b$ due to a stable field density in a long flux tube. The same
factor suggests our estimate to be maybe 20\% lower than the physical one
due to a higher quark mass. 

\section{An ``order parameter'' for string breaking?}

There has been speculation based on an instanton calculation~\cite{dia:90} 
that an adjoint string in pure SU(3) would not break at all -- this 
calculation has recently been heavily criticized~\cite{fab:99}. Some people 
are also looking for
an order parameter for fundamental string breaking in full QCD that would 
distinguish between
quenched and unquenched lattice configurations (the measurement programs for 
both are identical). The potential $V_0(R)$ measured with Wilson loops should
flatten only for unquenched, but this hasn't been found to actually happen as 
discussed above. The variational approach has been critized for giving
a flattening ground state potential also in the quenched model. 

In the quenched case the variational matrix $C(R,T)$ turns out to be  
inconsistent; $U \ne 0$, allowing the extraction of $x$ relevant for the full 
theory, but no mixing appears in the energies of the transfer matrix. There 
is no energy gap between ground and first excited state for quenched. This 
inconsistency is a reflection of the non-unitary nature of the quenched
approximation when light quark degrees of freedom are explicitly introduced
in the correlators. Reflection positivity is lost and there is no QFT left. 
Thus we don't think there is a need for an ``order parameter''. 
However, such a parameter can be obtained by e.g. measuring the $Q\bar{Q}$ 
and $Q\bar{q}+q\bar{Q}$ energies separately 
and looking for a gap between them at the string breaking point.

\section{Excited strings}

In the breaking of a first excited state of the flux tube conservation of its
angular momentum forces one of the resulting mesons to have $L>0$ as in the 
selection rule mentioned above. In practise de-excitation into a 
$Q\bar{Q}+q\bar{q}$ state seems to be a much more relevant decay 
channel~\cite{mic:99b}. In this and higher-lying cases the energies and
mixing between hybrid $Q\bar{Q}$, two heavy-light $Q\bar{q}+q\bar{Q}$ and
$Q\bar{Q}+q\bar{q}$ states can be studied on the lattice with our techniques,
including a new method for calculating disconnected diagrams~\cite{mic:99c}.

\section*{Acknowledgement}
This work was done as part of the UKQCD collaboration with C. Michael (Liverpool University). 

\newcommand{\href}[2]{#2}\begingroup\raggedright\endgroup

\end{document}